\documentclass[conference]{IEEEtran}
\usepackage[utf8]{inputenc}
\usepackage[T1]{fontenc}
\IEEEoverridecommandlockouts
\usepackage{cite}
\usepackage{amsmath,amssymb,amsfonts}
\usepackage{graphicx}
\usepackage{textcomp}
\usepackage{xcolor}
\usepackage{url}
\usepackage{booktabs}
\usepackage{array}
\usepackage{multirow}
\def\BibTeX{{\rm B\kern-.05em{\sc i\kern-.025em b}\kern-.08em
    T\kern-.1667em\lower.7ex\hbox{E}\kern-.125emX}}

\begin{document}

\title{Privacy-Preserving Smart Surveillance with Cross-Dataset Violence Detection and Decentralized Evidence Governance}

\makeatletter
\newcommand{\linebreakand}{%
 \end{@IEEEauthorhalign}
  \hfill\mbox{}\par
  \mbox{}\hfill\begin{@IEEEauthorhalign}
}
\makeatother

\author{
\IEEEauthorblockN{1\textsuperscript{st} Hasan Coşkun}
\IEEEauthorblockA{
\textit{Kadir Has University}\\
Istanbul, Turkey \\
hasancoskun@stu.khas.edu.tr
}
\and
\IEEEauthorblockN{2\textsuperscript{nd} Furkan Çolhak}
\IEEEauthorblockA{
\textit{Kadir Has University}\\
Istanbul, Turkey \\
furkancolhak@stu.khas.edu.tr
}
\and
\IEEEauthorblockN{3\textsuperscript{rd} Andrea Kulakov}
\IEEEauthorblockA{
\textit{Faculty of Computer Science and Engineering}\\
\textit{Ss. Cyril and Methodius University in Skopje}\\
Skopje, North Macedonia \\
andrea.kulakov@finki.ukim.mk
}

\linebreakand

\IEEEauthorblockN{4\textsuperscript{th} Vesna Dimitrova}
\IEEEauthorblockA{
\textit{Faculty of Computer Science and Engineering}\\
\textit{Ss. Cyril and Methodius University in Skopje}\\
Skopje, North Macedonia \\
vesna.dimitrova@finki.ukim.mk
}
}

\maketitle

\begin{abstract}
AI-enabled surveillance can accelerate public-safety response, yet most systems still leave recorded evidence under centralized administrative control. This paper proposes a privacy-preserving smart surveillance framework that separates incident detection from evidence disclosure. A lightweight MobileNetV2-based video classifier detects violent clips, while each recorded incident segment is immediately encrypted and made accessible only through threshold-based approval. The decryption key is split with Shamir's Secret Sharing, member shares are protected with public-key cryptography, and voting is supported by time-limited tokens, two-factor authentication, signatures, and audit logs. This study evaluates MobileNetV2+LSTM, MobileNetV2+BiLSTM, and MobileNetV2+temporal CNN heads on SCVD, RWF-2000, and Real-Life Violence Situations under seven in-domain and cross-dataset scenarios. The best all-source model, MobileNetV2+BiLSTM, reaches \textbf{93.5\%} test accuracy and ROC-AUC \textbf{0.980\%} on the merged held-out set, while lower RWF-2000 slice performance confirms persistent dataset shift.
\end{abstract}


\section{Introduction}

Smart-city surveillance systems increasingly combine camera networks with AI-based video analytics \cite{veena2025systematic}. Automated violence detection is attractive because it can reduce response latency and help preserve time-sensitive evidence \cite{nwokonkwo2024enhancing}. However, the same systems also create privacy and accountability risks: once footage is recorded, access is often controlled by a single operator, administrator, or institution \cite{richards2012dangers}. These concerns are consistent with broader public attitudes toward data governance; a Pew Research Center survey reports that 71\% of U.S. adults are concerned about how the government uses collected data, while 79\% feel they have little or no control over such data use~\cite{pew2023dataprivacy}. Accuracy alone therefore does not make an intelligent surveillance system trustworthy.

This paper addresses the gap between detection and evidence governance. We propose a framework in which a violence detector acts only as an incident trigger. After detection, the relevant video segment is encrypted immediately; the key is split among authorized members; and plaintext evidence can be recovered only after a threshold of authenticated approvals. The design follows privacy-by-design principles \cite{cavoukian2011privacy} and targets the specific risk of unilateral disclosure of sensitive public or semi-public footage.

\begin{table*}[t]
\centering
\scriptsize
\caption{Summary of selected violence-detection studies.}
\label{tab:selected_studies}
\begin{tabular}{@{}p{2.2cm}p{3.4cm}p{2.8cm}p{2.4cm}p{1.4cm}p{3.0cm}@{}}
\toprule
\textbf{Study} & \textbf{Approach} & \textbf{Dataset(s)} & \textbf{Reported Performance} & \textbf{Alert} & \textbf{Privacy / Governance} \\
\midrule
Ullah et al. \cite{ullah2021ai} & Edge-cloud YOLOv3-tiny with ConvLSTM/GRU. & Surveillance Fight, RWF-2000, Hockey Fight. & Acc. up to 88.2\% on RWF-2000. & Yes & Privacy motivation, but no threshold access. \\
Suba et al. \cite{suba2022violence} & Lightweight 3D ConvNet with MobileNetV2/ResNet variants. & Real-Life Violence, UBI-Fights, UCF-Crime. & F1 up to 0.97. & Yes & Not discussed. \\
Gao \cite{gao2023yolo} & YOLO-based violence detection. & Custom image dataset. & mAP about 91\%. & Mentioned & Not discussed. \\
Nivedita et al. \cite{nivedita2023real} & MobileNetV2 with LSTM/BiLSTM and YOLOv7. & Custom and public data. & Acc. 0.88; AP 0.87. & Yes & Not discussed. \\
Akole et al. \cite{akole2023real} & MobileNetV2, YOLO, HOG, optical flow. & Real-Life Violence Situations. & Acc. about 90\%. & Yes & Not discussed. \\
Jaafar et al. \cite{jaafar2024detection} & CNN-RNN and video transformers. & Surveillance and YouTube fight data. & Best F1 86.8\%. & Yes & Not discussed. \\
Khan et al. \cite{khan2024vdnet} & Edge-vision surveillance architecture. & Hockey, Movies, RWF-2000, Surveillance Fight. & Acc. up to 99.0\% on selected sets. & Yes & Edge locality, no decentralized evidence control. \\
Thomas and Balamurugan \cite{thomas2024real} & MobileNetV2 with cloud Firestore. & Custom dataset. & Acc. 96\%. & Yes & Cloud security mentioned; no voting/key sharing. \\
Trinh et al. \cite{trinh2024violence} & VGG19 features with ConvLSTM2D. & Hockey, Movies, Violent Flow. & Acc. up to 99.2\%. & Mentioned & Not discussed. \\
Wagh et al. \cite{wagh2024violence} & MobileNetV2 with cloud notification. & Real-Life Violence. & Acc. 95.76\%. & Yes & Storage security mentioned; centralized access remains. \\
Prabha et al. \cite{prabha2025hybrid} & MobileNetV2 with ConvLSTM. & Kaggle violence data. & Acc. 96\%. & Yes & Not discussed. \\
Jain et al. \cite{jain2025towards} & MobileNetV2 with temporal analysis and cloud notification. & Kaggle dataset. & Acc. 96\%. & Yes & Firestore/Telegram security, no threshold governance. \\
\bottomrule
\end{tabular}
\end{table*}

The machine learning component is evaluated on three public datasets: SCVD, RWF-2000, and Real-Life Violence Situations. We compare MobileNetV2+LSTM, MobileNetV2+BiLSTM, and MobileNetV2+temporal CNN variants under seven scenarios that separate in-domain performance from cross-dataset transfer. The experimental design is intentionally conservative: held-out and cross-domain test sets are not used for checkpoint selection, and the E7 all-source condition is also reported by per-source slices.

The main contributions of this paper are as follows:
\begin{itemize}
    \item We propose an end-to-end privacy-preserving smart surveillance framework that integrates violence detection, immediate evidence encryption, distributed key management, threshold-based voting, and auditability.
    \item We evaluate lightweight MobileNetV2-based LSTM, BiLSTM, and temporal CNN architectures across SCVD, RWF-2000, and Real-Life Violence Situations using seven in-domain and cross-dataset scenarios.
    \item We introduce a decentralized evidence-governance workflow in which recorded footage can be decrypted only after authenticated multi-party approval.
    \item We release the implementation and experimental setup as a reproducibility artifact to support independent inspection and extension.
\end{itemize}

\section{Related Work}
\label{sec:related_work}

Recent violence-detection research has moved from hand-crafted motion features toward deep CNN, recurrent, ConvLSTM, YOLO-based, and transformer-based video models. Lightweight backbones such as MobileNetV2 remain common in real-time or edge-oriented systems because they reduce inference cost while preserving useful visual features \cite{suba2022violence,thomas2024real,wagh2024violence}. RWF-2000, Real-Life Violence Situations, UCF-Crime, UBI-Fights, and SCVD are frequently used benchmarks, but many studies still emphasize in-domain accuracy rather than robustness to unseen camera distributions \cite{cheng2021rwf,kaggle_violence,sultani2018real,aremu2024ssivd}.

The reviewed literature also shows a governance gap. Most systems include some alerting mechanism, such as alarms, messaging, dashboards, or cloud notification, but few specify how recorded evidence is encrypted, who may authorize access, or how misuse by privileged actors is prevented \cite{ullah2021ai,khan2024vdnet,wagh2024violence}. Cloud storage security or edge processing can reduce some exposure, yet they do not by themselves prevent unilateral access to footage \cite{fitwi2021privacy}. This work therefore combines cross-dataset violence detection with cryptographic evidence governance.

Table~\ref{tab:selected_studies} motivates the central design choice of this paper. Prior work demonstrates that compact deep models can detect violence with promising accuracy, but detection and alerting usually end the technical contribution. In conventional AI-enabled surveillance pipelines, the ability to detect an incident is therefore not matched by an equally strong mechanism for governing access to the resulting video evidence. Once a clip is recorded, it may still be stored and accessed through centralized administrative control, leaving unresolved risks of unauthorized viewing, insider misuse, or unilateral disclosure.

In contrast, the proposed framework separates incident detection from evidence disclosure. The violence detector continuously analyzes video streams and triggers recording only when a violent event is detected. The recorded segment is then encrypted immediately, and the decryption authority is distributed among authorized members rather than retained by a single operator or administrator. Plaintext evidence can be recovered only after a threshold-based voting process reconstructs the decryption key. This design positions the proposed system as both a violence-detection framework and a privacy-preserving evidence-governance mechanism.
\section{Proposed Framework}
\label{sec:system_overview}

The proposed framework separates \textit{incident detection} from \textit{evidence disclosure}. A video stream is analyzed by a violence detector; when the predicted probability exceeds a configured threshold, the system records the relevant segment. This segment is not stored as accessible plaintext. It is encrypted immediately with an incident-specific symmetric key, and the key is distributed among authorized members through threshold secret sharing. Plaintext evidence can be recovered only if at least $k$ of $n$ members approve access. Figure~\ref{fig:system_overview_placeholder} marks the intended two-column architecture figure.

\begin{figure*}[t]
\centering
\includegraphics[width=\textwidth]{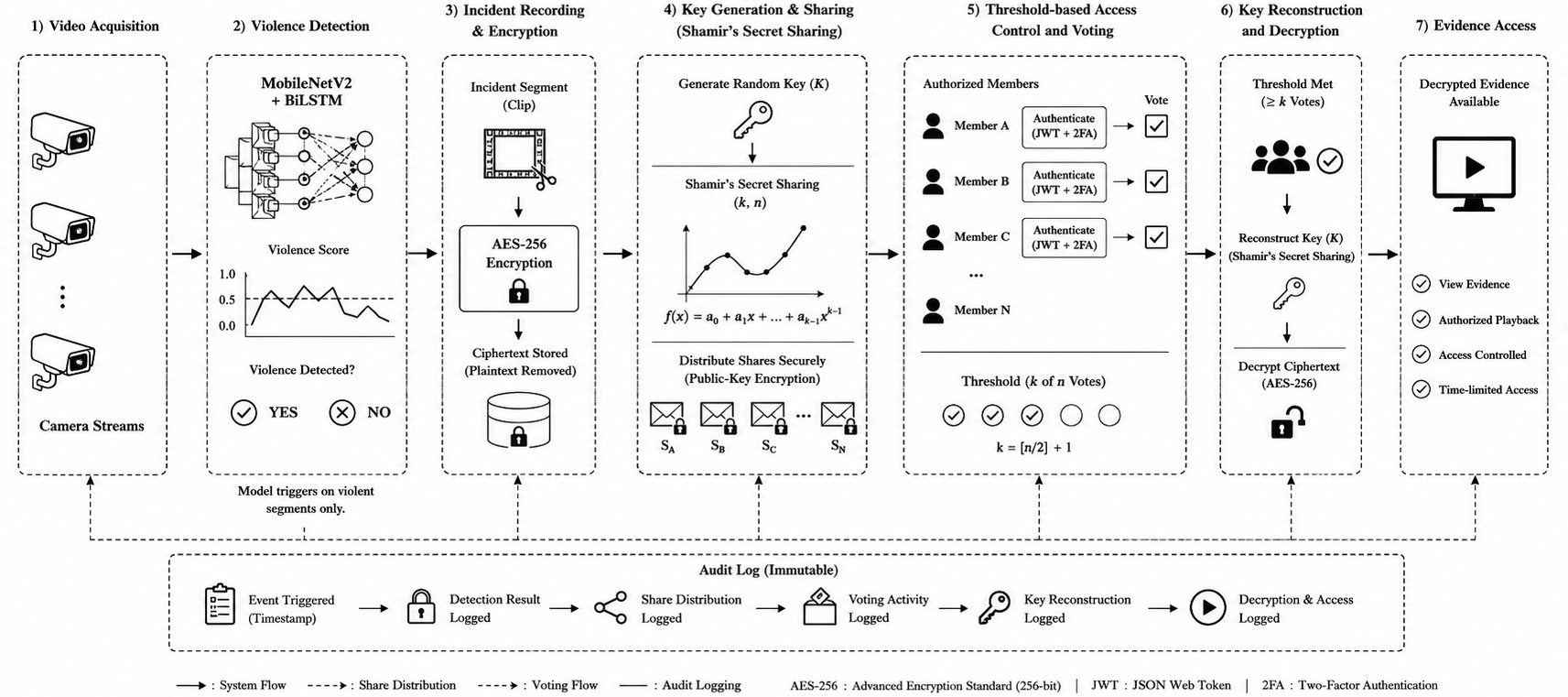}
\caption{Overview of the proposed privacy-preserving smart surveillance framework.}
\label{fig:system_overview_placeholder}
\end{figure*}

The workflow has four layers. The detection layer buffers frames, applies the trained MobileNetV2-based classifier, and emits an incident event. The evidence layer encrypts the recorded clip and removes the transient plaintext copy. The key-management layer splits the decryption key using Shamir's Secret Sharing \cite{shamir1979share}. The governance layer creates a voting session, authenticates members, records signed votes, checks the threshold, reconstructs the key only after approval, and writes audit records for accountability.

For a voting body of $n$ members, the default approval threshold is a strict majority:
\begin{equation}
k = \left\lfloor \frac{n}{2} \right\rfloor + 1.
\label{eq:threshold}
\end{equation}
This does not remove the need for legal and institutional rules \cite{GDPR2016}, but it changes the technical trust model. A single administrator, database operator, or member should not be able to decrypt evidence alone. The model output is therefore used to preserve evidence, while human access remains a cryptographically mediated collective decision.

\section{Detection Methodology}
\label{sec:ml_methodology}

Violence detection is formulated as binary video classification:
\begin{equation}
    y \in \{\textit{normal}, \textit{violence}\}.
\end{equation}
The classifier estimates $p_{\theta}(y=\textit{violence}\mid x)$ for a short clip $x$. A positive prediction triggers evidence preservation, but it does not authorize evidence disclosure.

The experiments use SCVD, RWF-2000, and Real-Life Violence Situations. SCVD provides CCTV-style smart-city footage; its \textit{weaponized} class is excluded to maintain a common binary label space. RWF-2000 contains balanced real-world fight and non-fight surveillance-like clips \cite{cheng2021rwf}. Real-Life Violence Situations provides diverse violent and non-violent clips from mixed real-world sources \cite{kaggle_violence}. The final manifest contains 6105 labeled clips.

Each clip is represented by 16 uniformly sampled RGB frames resized to $224 \times 224$ and normalized with the MobileNetV2 preprocessing function. Uniform sampling preserves coarse temporal coverage while keeping the input size suitable for lightweight deployment. All architectures share an ImageNet-pretrained MobileNetV2 frame encoder and differ only in the temporal head: a unidirectional LSTM \cite{hochreiter1997long}, a bidirectional LSTM \cite{schuster1997bidirectional}, or a temporal CNN/pooling head.

Models are trained with Adam at an initial learning rate of $10^{-5}$, batch size 4, and a maximum of 100 epochs. Early stopping and checkpoint selection use validation ROC-AUC with patience 5, and cross-domain tests are evaluated only after model selection. This prevents hidden leakage from the cross-dataset evaluation sets into training decisions.

\section{Evidence Governance Protocol}
\label{sec:evidence_governance}

The governance protocol protects recorded evidence against unauthorized viewing, rogue administrators, database compromise, replayed voting links, vote manipulation, and single-member key exposure. It assumes standard cryptographic primitives are implemented by maintained libraries and that fewer than $k$ voting members collude maliciously \cite{shamir1979share}. If $k$ or more authorized members approve access, the system treats the result as a governance decision rather than a cryptographic failure.

For each incident, the recorded plaintext video $P$ is encrypted with a fresh AES-256 key $K$ and initialization vector $IV$ \cite{NIST2001AES,NIST2001modes}:
\begin{equation}
    C = \mathrm{AES\mbox{-}CBC}_{K,IV}(P).
    \label{eq:aes_encryption}
\end{equation}
The ciphertext and non-sensitive metadata are stored, while $K$ is never retained as a directly accessible value. Instead, $K$ is interpreted as a secret $s$ in the prime field $\mathbb{Z}_p$ defined next, and split with a degree-$(k-1)$ polynomial:
\begin{equation}
    f(x) = s + a_1x + \cdots + a_{k-1}x^{k-1} \pmod{p}.
    \label{eq:sss_polynomial}
\end{equation}
In our prototype, Shamir's scheme is instantiated in the prime field $\mathbb{Z}_p$ with $p = 2^{521}-1$ (a Mersenne prime). The AES-256 key $K$ is interpreted as an integer secret $s<p$ represented in big-endian byte order, consistent with~\eqref{eq:sss_polynomial} and~\cite{shamir1979share}.

Each member receives a share $(i,f(i))$, protected with that member's RSA-OAEP public key \cite{RSA2012PKCS1}. Once at least $k$ valid approval shares are collected, the key is reconstructed by Lagrange interpolation.

Voting links are bound to a session and member through signed JWT claims \cite{RFC7519}; vote submission also requires a TOTP code \cite{RFC6238}. Each vote is recorded with a digital signature, and database constraints prevent duplicate votes for the same member-session pair. Audit logs record incident creation, encryption, share generation, member notification, vote submission, threshold evaluation, key reconstruction, and evidence decryption. The result is an evidence workflow in which detection creates an encrypted object, while disclosure requires authenticated and auditable threshold approval.

\section{Implementation}
\label{sec:implementation}

The prototype is organized as a modular Python service. OpenCV handles webcam, RTSP/IP camera, and file-based ingestion. A model adapter performs frame preprocessing, inference, class mapping, and thresholding, allowing trained Keras or exported models to be swapped through configuration rather than by changing the incident workflow.

For reproducibility, the full implementation and experimental setup are made available in the project repository\footnote{\url{https://github.com/furkancolhak/democratic-surveillance-system}}. The repository contains the privacy-preserving smart surveillance framework described in this paper, including the violence detection module, AES-256-CBC encryption service, Shamir's Secret Sharing implementation, community voting portal with TOTP two-factor authentication and RSA digital signatures, PostgreSQL database schema, Flask API, Docker deployment configuration, and Nginx reverse-proxy setup. This artifact is intended to support independent inspection, replication of the system workflow, and extension of the experimental pipeline.

The web and governance layer is implemented with Flask served by Gunicorn. PostgreSQL and SQLAlchemy store administrators, members, incidents, encrypted shares, votes, and audit logs. Composite uniqueness constraints prevent duplicate member votes, foreign keys preserve session relationships, and transactions keep multi-step voting operations consistent. Cryptographic services implement evidence encryption, key splitting, share protection, vote signing, and secure handling of member-side secrets. Deployment uses Docker Compose with a PostgreSQL service, an application service, and an Nginx reverse proxy for TLS termination, rate limiting, security headers, and request forwarding.

This implementation is meant to demonstrate feasibility rather than claim production hardening. A deployment-grade version should replace CBC-only encryption with authenticated encryption, use stronger key-management infrastructure such as HSM or KMS-backed keys, and validate operational latency on the target camera network.

\section{Experimental Protocol}
\label{sec:experimental_evaluation}

Let $S$, $R$, and $L$ denote SCVD, RWF-2000, and Real-Life Violence Situations. Seven scenarios are used. E1--E3 train on one source and test on both its held-out split and the two withheld sources. E4--E6 train on two-source unions and test on the held-out union plus the withheld third source. E7 trains on all three sources and reports a merged held-out set with source-specific slices. The E7 test pool contains 918 clips: 321 from SCVD, 297 from RWF-2000, and 300 from Real-Life Violence Situations.

The protocol distinguishes in-domain testing from cross-domain testing. In-domain clips come from sources represented during training and validation; cross-domain clips come from sources fully withheld from training and validation. This distinction is essential because models that perform well within one dataset may not generalize to different visual domains, a well-known issue in computer vision dataset bias and cross-dataset evaluation~\cite{torralba2011datasetbias,tommasi2017datasetbias}. In surveillance violence detection, this concern is especially relevant because datasets differ in camera viewpoint, scene structure, lighting, motion patterns, compression artifacts, and annotation conventions~\cite{cheng2021rwf}.

Results are reported as accuracy and ROC-AUC:
\begin{equation}
    \mathrm{Accuracy} =
    \frac{TP + TN}{TP + TN + FP + FN}.
\end{equation}
ROC-AUC is computed as the area under the receiver operating characteristic curve:
\begin{equation}
    \mathrm{ROC\mbox{-}AUC} =
    \int_{0}^{1} \mathrm{TPR}(\mathrm{FPR})\,d\mathrm{FPR}.
\end{equation}
Here, $\mathrm{TPR}=TP/(TP+FN)$ and $\mathrm{FPR}=FP/(FP+TN)$. ROC-AUC is included because it measures ranking quality across thresholds and is less tied to a single operating point than accuracy. No reported test set is used for early stopping, learning-rate scheduling, or model selection.

\section{Results}
\label{sec:results}

Results are reported as accuracy / ROC-AUC. Three patterns are consistent across the E1--E7 protocol. First, single-source training can produce excellent in-domain scores but weaker cross-domain transfer. Second, adding a second source improves diversity but does not remove the difficulty of a fully withheld dataset. Third, all-source training gives the strongest pooled held-out result, while per-source slices still reveal dataset-specific weakness.

\begin{table*}[t]
\centering
\scriptsize
\caption{Single-source training results. Values are accuracy / ROC-AUC.}
\label{tab:single_source_results}
\begin{tabular}{@{}llccc@{}}
\toprule
\textbf{Scenario} & \textbf{Model} & \textbf{In-Domain Hold-Out} & \textbf{Cross-Test 1} & \textbf{Cross-Test 2} \\
\midrule
E1: train SCVD & MobileNetV2+LSTM & \textbf{0.988 / 1.000} & RWF: 0.670 / 0.734 & RLV: 0.690 / 0.771 \\
E1: train SCVD & MobileNetV2+BiLSTM & \textbf{0.994 / 1.000} & RWF: 0.636 / 0.711 & RLV: 0.627 / 0.843 \\
E1: train SCVD & MobileNetV2+Temporal CNN & \textbf{0.994 / 1.000} & RWF: 0.650 / 0.705 & RLV: 0.700 / 0.793 \\
\midrule
E2: train RWF & MobileNetV2+LSTM & \textbf{0.818 / 0.894} & SCVD: 0.660 / 0.702 & RLV: 0.720 / 0.849 \\
E2: train RWF & MobileNetV2+BiLSTM & \textbf{0.828 / 0.890} & SCVD: 0.679 / 0.739 & RLV: 0.697 / \textbf{0.875} \\
E2: train RWF & MobileNetV2+Temporal CNN & \textbf{0.785 / 0.897} & SCVD: 0.676 / 0.742 & RLV: 0.657 / 0.853 \\
\midrule
E3: train RLV & MobileNetV2+LSTM & \textbf{0.963 / 0.993} & SCVD: 0.598 / 0.637 & RWF: 0.620 / 0.710 \\
E3: train RLV & MobileNetV2+BiLSTM & \textbf{0.967 / 0.995} & SCVD: 0.632 / 0.647 & RWF: 0.609 / 0.707 \\
E3: train RLV & MobileNetV2+Temporal CNN & \textbf{0.967 / 0.993} & SCVD: 0.583 / 0.633 & RWF: 0.596 / 0.705 \\
\bottomrule
\end{tabular}
\end{table*}

Table~\ref{tab:single_source_results} shows that SCVD-only and RLV-only models nearly saturate their own held-out splits, yet transfer poorly to other sources. RWF-only training is more moderate in-domain and also transfers unevenly. These results indicate strong dataset dependence and justify the two-source and all-source experiments.

\begin{table*}[t]
\centering
\scriptsize
\caption{Two-source training results. Values are accuracy / ROC-AUC.}
\label{tab:two_source_results}
\begin{tabular}{@{}llcc@{}}
\toprule
\textbf{Scenario} & \textbf{Model} & \textbf{Held Union} & \textbf{Withheld Source Cross-Test} \\
\midrule
E4: train SCVD+RWF & MobileNetV2+LSTM & 0.892 / 0.966 & RLV: 0.793 / 0.869 \\
E4: train SCVD+RWF & MobileNetV2+BiLSTM & 0.896 / 0.967 & RLV: 0.780 / 0.874 \\
E4: train SCVD+RWF & MobileNetV2+Temporal CNN & 0.901 / 0.968 & RLV: 0.767 / 0.874 \\
\midrule
E5: train SCVD+RLV & MobileNetV2+LSTM & 0.982 / 0.996 & RWF: 0.677 / 0.774 \\
E5: train SCVD+RLV & MobileNetV2+BiLSTM & 0.982 / \textbf{0.998} & RWF: 0.667 / 0.766 \\
E5: train SCVD+RLV & MobileNetV2+Temporal CNN & 0.982 / 0.997 & RWF: 0.670 / 0.753 \\
\midrule
E6: train RWF+RLV & MobileNetV2+LSTM & 0.888 / 0.962 & SCVD: 0.689 / 0.745 \\
E6: train RWF+RLV & MobileNetV2+BiLSTM & 0.901 / 0.963 & SCVD: 0.738 / 0.780 \\
E6: train RWF+RLV & MobileNetV2+Temporal CNN & 0.889 / 0.959 & SCVD: 0.704 / 0.770 \\
\bottomrule
\end{tabular}
\end{table*}

Table~\ref{tab:two_source_results} shows that combining sources improves held-union performance, but the withheld source can remain difficult. The RWF cross-test in E5 is the clearest example: despite near-perfect SCVD+RLV held-union scores, RWF transfer remains around $0.67$ accuracy.

\begin{table*}[t]
\centering
\scriptsize
\caption{E7 all-source training results and source-specific slices. Values are accuracy / ROC-AUC.}
\label{tab:e7_results}
\begin{tabular}{@{}lcccc@{}}
\toprule
\textbf{Model} & \textbf{Merged} & \textbf{SCVD-only} & \textbf{RWF-only} & \textbf{RLV-only} \\
\midrule
MobileNetV2+LSTM & 0.922 / 0.978 & 0.978 / 0.998 & 0.811 / 0.908 & 0.970 / 0.995 \\
MobileNetV2+BiLSTM & \textbf{0.935 / 0.980} & \textbf{0.994 / 1.000} & \textbf{0.828 / 0.901} & \textbf{0.977 / 0.997} \\
MobileNetV2+Temporal CNN & 0.923 / 0.979 & 0.981 / 0.999 & 0.818 / 0.906 & 0.963 / 0.994 \\
\bottomrule
\end{tabular}
\end{table*}

Table~\ref{tab:e7_results} reports the strongest training condition: all three datasets are used for training and validation, and the merged held-out set contains SCVD, RWF-2000, and RLV clips. MobileNetV2+BiLSTM achieves the best merged result, $0.935$ accuracy and $0.980$ ROC-AUC. The temporal CNN head is close in ROC-AUC, suggesting that lightweight temporal aggregation may be attractive when inference cost matters.

The source slices are more important than the pooled score alone. SCVD and RLV are classified with very high accuracy, but the RWF-only slice remains lower for every architecture. This suggests that RWF-2000 contains visual or temporal patterns not fully absorbed by the all-source model. Future improvements should therefore prioritize source-balanced training, RWF-aware augmentation, and deployment-specific calibration rather than relying only on a larger pooled score.

\section{Discussion}
\label{sec:discussion}

The results support the paper's main architectural claim: violence detection should trigger protected preservation, not automatic evidence disclosure. Even the strongest E7 model remains source-dependent, especially on RWF-2000. False positives may create unnecessary incident records, while false negatives may fail to preserve evidence; immediate encryption and threshold-governed access reduce the privacy impact of false positives by preventing automatic human exposure.

The cryptographic layer reduces insider risk by replacing unilateral administrative access with encrypted evidence, Shamir-based key splitting, protected shares, and reconstruction only after at least $k$ authenticated approvals~\cite{dworkin2001sp80038a,shamir1979share,RSA2012PKCS1}. Thus, database-only compromise should expose encrypted records rather than usable evidence, and a single compromised member should not reconstruct the key under the threshold assumption~\cite{shamir1979share}. This aligns with least-privilege and separation-of-duties principles~\cite{nist80053r5,nist800171sod}.

The design remains bounded: threshold secret sharing cannot prevent collusion by $k$ or more members~\cite{shamir1979share}, TOTP is not phishing-resistant~\cite{nist80063b}, audit logs require separate protection~\cite{nist80092}, and higher thresholds improve collusion resistance at the cost of response latency. Overall, the framework is a privacy-preserving evidence-management pattern for AI surveillance: no model prediction or single privileged account should be enough to disclose sensitive footage.
\section{Limitations}
\label{sec:limitations}

This study has four main limitations. First, three public datasets cannot fully represent real camera-network diversity, so deployment requires local validation and calibration. Second, the task is binary, while real systems may need multi-class detection for weapons, vandalism, medical emergencies, panic, or accidents. Third, clip-level MobileNetV2 evaluation does not capture all real-time constraints such as stream instability, multi-camera scheduling, and latency. Fourth, the cryptographic prototype requires production hardening, including authenticated encryption, stronger key management, tamper-evident logs, legal review, retention rules, and transparent voting-member selection.
\section{Conclusion}
\label{sec:conclusion}
This paper presented a privacy-preserving smart surveillance framework that combines lightweight violence detection with encrypted, threshold-governed evidence access. The core principle is separation: machine learning detects and preserves candidate incidents, while cryptographic governance determines whether the resulting evidence may be disclosed.

The proposed workflow addresses a key weakness of alert-centric surveillance pipelines, where recorded footage often remains under centralized administrative control. By encrypting each incident segment immediately, splitting the decryption key among authorized members, and requiring authenticated threshold approval, the system makes unilateral disclosure technically harder and evidence access more auditable.

The evaluation across SCVD, RWF-2000, and Real-Life Violence Situations shows both the promise and limits of the detection layer. The all-source MobileNetV2+BiLSTM model achieved the best merged held-out result, with $93.5$ test accuracy and ROC-AUC $0.980$, but the weaker RWF-2000 slice confirms that dataset shift remains a deployment risk. Thus, high pooled performance should not be treated as automatic authority for evidence disclosure.

The released implementation demonstrates that privacy-preserving evidence governance can be integrated into an operational surveillance stack, including detection, encryption, secret sharing, voting, database, API, and deployment components. Future work should improve cross-source robustness, extend the task beyond binary violence detection, replace CBC-only encryption with authenticated encryption, strengthen key management and authentication, and make audit logs tamper-evident. Overall, intelligent surveillance should be evaluated not only by what it detects, but also by how responsibly it governs the evidence it creates.

\section{Acknowledgments}\label{sec7}
This work was supported partially by the European Union in the framework of ERASMUS MUNDUS, Project CyberMACS (Project \#101082683) (\url{https://cybermacs.eu}).\\

\bibliographystyle{ieeetr}
\bibliography{sn-bibliography}

\end{document}